\documentclass[fleqn,twoside]{article}
\usepackage{espcrc2}

\newcommand{\AmS}{{\protect\the\textfont2
  A\kern-.1667em\lower.5ex\hbox{M}\kern-.125emS}}

\title{A rigorous bound on quark distributions in the nucleon}

\author{E. Di Salvo \address[MCSD]{Dipartimento di Fisica
          and INFN - sez. Genova \\ 
        Via Dodecaneso 33, 16143 Genova, Italy}}
        
\begin{document}

\begin{abstract}
I deduce an inequality between the helicity and the transversity distribution
of a quark in a nucleon, at small energy scales. Then I establish, thanks to 
the positivity constraint, a rigorous bound on longitudinally polarized 
valence quark densities, which finds nontrivial applications to $d$-quarks. 
This, in turn, implies a bound for
the distributions of the longitudinally polarized sea, which is probably
not $SU(3)$-symmetric. Some model predictions and parametrizations of quark 
distributions are examined in the light of these results.   
\vspace{1pc}
\end{abstract}

\maketitle

\section{INTRODUCTION}

$~~$ The leading twist distributions of quarks and antiquarks 
inside the nucleon are not completely known. Indeed, the longitudinally 
polarized sea distributions are only poorly 
known\cite{lss,bt,rm}, while the transversely polarized densities are 
unknown at all. In this situation it is quite important to establish 
rigorous inequalities, which are very useful in best fits\cite{lss,bt} to data. 
As an example, I recall the famous Soffer inequality\cite{so}, which is a 
consequence of the positivity constraint. Here I deduce, as an application of 
that inequality, a rigorous bound concerning unpolarized and longitudinally 
polarized valence quark distributions. The 
talk is organized as follows. First of all (sect. 2), starting from the 
Soffer inequality, from the Melosh-Wigner rotation\cite{ma,ss,co} and from 
general considerations on evolution equations, I prove bound
(\ref{ineq3p}) for any $Q^2$-value, $-Q^2$ being the modulus square of the 
four-momentum of the probe ({\it e. g.}, a virtual photon). Secondly (sect. 3),
I consider applications 
to {\it down} quark distributions and to the first moments of polarized sea 
distributions. Lastly, in sect. 4, I draw a short conclusion.

\section{A RIGOROUS BOUND}

\subsection{The Melosh-Wigner rotation}

First of all, I deduce, for very small values of $Q^2$, an inequality between 
the helicity and the transversity distributions, for which I adopt 
the notations by Mulders and Tangerman\cite{mt}.

In the nucleon rest frame, let 
$q_0^{\pm}({\bf p})$ be the probability density for a quark of
momentum ${\bf p}$ to have spin parallel (+) or antiparallel (-)  
to the proton spin. A boost parallel to the nucleon spin, and such 
that the nucleon momentum becomes much greater than the nucleon
rest mass $M$, produces a spin dilution in the transverse momentum 
(t.m.) polarized density, which turns out to coincide with the 
longitudinally polarized t.m. distribution, {\it i. e.},
\begin{equation}
g_{1}(x, {\bf p_{\perp}}) = \left[q_0^+({\bf p}) - q_0^-({\bf 
p})\right] cos\theta_{M}. \label{g1l} 
\end{equation}
Here $x$ = $(p_0+p_3)/\sqrt{2}M$, ${\bf p_{\perp}}$ = ${\bf 
p}-p_3{\bf k}$, ${\bf k}$  is a unit vector in the direction of the 
boost, $p_3$ = ${\bf p}\cdot {\bf k}$, $p_0^2$ = 
$m^2+p_3^2+{\bf p}_{\perp}^2$
and $m$ is the quark rest mass. Lastly $\theta_{M}$ is the 
Melosh-Wigner rotation angle\cite{ma,ss,co}, {\it i. e.},
\begin{equation}
\theta_{M} = arccos\left[\frac{(m+\sqrt{2}xM)^2-{\bf p}_{\perp}^2}
{(m+\sqrt{2}xM)^2+{\bf p}_{\perp}^2}\right]. \label{long}
\end{equation}

On the other hand, a boost from the nucleon rest system, 
similar to the previous one, but in a 
direction perpendicular to its spin, produces a less drastic spin dilution. 
Indeed, in this case the distribution results 
in the t.m. transversity, the Melosh-Wigner rotation giving\cite{ss}
\begin{equation}
h_{1}(x, {\bf p}_{\perp}) = \left[q_0^+({\bf p}) - q_0^-({\bf 
p})\right] D_{\perp}(\theta_{M},\phi). \label{tras}
\end{equation}
Here
\begin{equation}
D_{\perp}(\theta_{M},\phi) = cos^2\frac{\theta_{M}}{2} + 
sin^2\frac{\theta_{M}}{2}(2sin^2\phi-1) \label{tras1}
\end{equation}
and $\phi$ is the azimuthal angle of ${\bf p_{\perp}}$ with respect to 
the plane perpendicular to the nucleon spin vector. Eqs. (\ref{g1l}) to 
(\ref{tras1}) imply
\begin{equation}
\frac{h_1(x)}{g_1(x)}\geq 1, \label{ineq1}
\end{equation} 
where $g_1(x) = \int d^2p_{\perp} g_{1}(x, {\bf p_{\perp}}^2)$ and 
$h_1(x)$ is defined analogously. This 
inequality - which reduces to equality for a nonrelativistic bound 
state - holds at the starting point for QCD 
evolution\cite{pp,gr,jr,bdr}, {\it i. e.,} for very small values of $Q^2$
($<< 1$ $GeV^{2}$). This is confirmed by 
previous calculations, based on the constituent quark model\cite{ma,ss}, 
on the bag model\cite{jj,sv}, on light cone models\cite{sw} or on 
the chiral quark model\cite{mg} (see also ref. 12 for a review).

\subsection{The Soffer inequality}

At increasing $Q^2$, $h_1$ decreases much more rapidly than $g_1$, owing 
to a different evolution kernel. Therefore inequality (\ref{ineq1}) no longer 
holds true for sufficiently large values of $Q^2$. However, as I shall show 
in the next subsection, relation (\ref{ineq1}), together with the Soffer 
inequality, {\it i.e.,}
\begin{equation}
2|h_1(x)| \leq f_1(x) + g_1(x), \label{sof} 
\end{equation} 
implies an inequality which holds for any $Q^2$. 

Indeed, (\ref{ineq1}) and (\ref{sof}) yield
\begin{equation}
2|g_{1}(x)| \leq f_1(x) + g_{1}(x), \label{ineq2}
\end{equation}
which is nontrivial for negative values of $g_1(x)$; in this case one has
\begin{equation}
-3 g_1(x) \leq f_1(x). \label{ineq3}
\end{equation}
Since $f_1(x)$ is nonnegative, inequality (\ref{ineq3}) holds true 
for {\it any} value of $x$. It is interesting to compare this 
inequality with a result of the nonrelativistic SU(6) quark model. 
For a {\it valence} $d$-quark in the nucleon, this model predicts 
\begin{equation}
g^d_{1v}(x) = -\frac{1}{3} f^d_{1v}(x).  \label{eq3}
\end{equation}
As I have 
shown above, in a relativistic bound state with spinning constituents
one has to take into account the Melosh-Wigner rotation, which produces 
a spin dilution (see also\cite{ma,ck}). Therefore eq. (\ref{eq3}) is 
to be replaced by
\begin{equation}
0 < -g^d_{1v} (x) \leq \frac{1}{3} f^d_{1v}(x), \label{ineq33}
\end{equation}
whose upper bound is a particular case of (\ref{ineq3}).

\subsection{Evolution equations}

Now I show that, under rather general assumptions, inequality 
(\ref{ineq3}), if referred to {\it valence} quarks, holds true for any $Q^2$
$>$ $Q_0^2$, with $Q_0^2$ $\leq$ 1 $GeV^2$. To this end, define the function
\begin{equation}
\phi(x,t) = f_{1v}(x,t)+3g_{1v}(x,t), ~~ \ t = ln 
\frac{Q^2}{Q^2_0},
\end{equation}
and set $\phi(x,0) > 0$. For $Q_0^2$ $\simeq$ 1 
$GeV^2$, it looks reasonable to assume, like Bourrely et al.\cite{bst},  
that $f_{1v}(x,t)$ and $g_{1v}(x,t)$ evolve 
according to the DGLAP equations at least for $Q^2 \geq Q^2_0$. Then, 
for $t$ $\geq$ $0$, the leading order (LO) QCD evolution equation reads
\begin{eqnarray}
\frac{d}{dt}\phi(x,t) &=& \frac{\alpha_s(t)}{2\pi} \int_x^1 \frac{dy}{y}
\phi(y,t) P(\frac{x}{y}), \label{ev} 
\\ 
P(z) &=& C_F \left(\frac{1+z^2}{1-z^2}\right)_+. 
\end{eqnarray}
This is a consequence of the fact that at LO $f_{1v}(x,t)$ and $g_{1v}(x,t)$ 
have the same evolution kernel. Since  eq. (\ref{ev}) preserves 
positivity\cite{bst}, at LO $\phi(x,t)$ is positive for any positive value of 
$t$. The effects of the next-to-leading order (NLO) approximation - which makes
the quark densities scheme dependent and introduces a complication in the 
evolution of $\phi$ - are completely screened by the LO term\cite{blt} and 
therefore do not affect the result\cite{ar,blt}. 

Now, if I push $Q_0^2$ down to very small values 
($<< 1 ~ GeV^2$), a nonperturbative evolution of the type described in 
refs.\cite{sw,bf,dis3} - based essentially on the chiral quark 
model\cite{mg} - could be assumed. In this case the elementary process 
determining the evolution is 
\begin{equation}
q \to \pi q', \label{elpr}
\end{equation}
where $q$ is a quark and $\pi$ a pion. Considerations analogous to the DGLAP 
equation can be done in the framework of this model. In fact the LO evolution 
is described by an equation similar to (\ref{ev}), {\it i. e.},
\begin{equation}
\frac{d}{dt}\phi(x,t) = \frac{g_{\pi}^2}{8\pi^2} \int_x^1 \frac{dy}{y}
\phi(y,t) P'(\frac{x}{y}). \label{evp} 
\end{equation}
Here $g_{\pi}$  is the pion-quark coupling constant and $P'(z)$ the 
splitting function for the process (\ref{elpr}). Since $P'(z)$ again 
preserves positivity, and $g_{\pi}$ is small enough to assure the  
screening of NLO effects by the LO term, the model leads to conclusions 
similar to those drawn in the framework of perturbative QCD.  

It is important to realize that, in the case of nonperturbative
evolution, the result we have found does not depend on the specific 
model assumed. Indeed, the same reasoning holds equally true for any
evolution mechanism of the type
\begin{equation}
q \to B q', \label{elsec}
\end{equation}
where $q$ is a quark and $B$ a boson - not necessarily a gluon or a 
pion -, provided the probability of such an elementary process 

~~i) is sufficiently small,

~ii) preserves positivity, 

iii) is helicity independent at LO.

But the third requirement is a necessary consequence of helicity 
conservation and implies that $f_{1v}(x,t)$ and $g_{1v}(x,t)$ have 
the same LO evolution kernel. 
On the other hand, any realistic process of the 
type (\ref{elsec}) satisfies conditions i) and ii). Therefore I conclude  
that inequality (\ref{ineq3}) is true under quite general assumptions, 
and practically for any $Q^2$, since $Q_0^2$ may assume very small 
values. This constitutes a nontrivial bound to the 
parametrizations of the quark distribution functions. In fact, it implies, 
together with the positivity constraint, 
\begin{equation}
-1/3 f_{1v}(x) \leq g_{1v}(x) \leq f_{1v}(x), \label{ineq3p}
\end{equation}
which is stronger than the inequality
\begin{equation}
|g_1(x)| \leq f_1(x),
\end{equation}
usually taken into account in the fits to data of polarized deep inelastic 
scattering\cite{lss}.

\section{APPLICATIONS}

Now I compare bound (\ref{ineq3p}) with models and parametrizations of 
quark distributions, especially as to the {\it down} flavor. For example, the 
predictions of some models, like the constituent quark model\cite{is} 
and the Carlitz-Kaur model\cite{ck}, fulfil inequality (\ref{ineq3p}).
On the contrary, in a best fit to semi-inclusive data\cite{l1}, the ratio 
$g_1^d(x)/f_1^d(x)$ does not satisfy this inequality for $x > 0.1$; 
this casts some doubts on that fit, since for sufficiently large $x$ 
($x$ $>$ $x_0$, with $x_0$ $\sim$ $0.2$)
the quark distributions derive their contributions essentially from 
valence quarks.

Integrating over $x$ from 0 to 1 all the terms which appear in the 
inequality (\ref{ineq3p}) yields
\begin{equation}
-n_q/3  < \Delta q_v \leq n_q. \label{ineq4}
\end{equation} 
Here $\Delta q_v$ is the first moment of $g_{1v}(x)$ and $n_q$ the 
valence number, $n_q$ = 2 for $u$-quarks and 1 for $d$-quarks.  
The lower bound (\ref{ineq4}) is not respected by the $d$-quark 
parametrizations deduced from the best fits in the 
literature\cite{lss,bt}. Leader, Sidorov and Stamenov\cite{lss}, who  
assume an $SU(3)$-symmetric polarized sea, find, at $Q^2$ = 1 $GeV^2$,
$-\Delta d_v$ = 0.339-0.341. On the other hand, the 
best fits by Bartelski and Tatur\cite{bt}, who do not assume any constraint 
for the sea, result, at the same $Q^2$-value, in $-\Delta d_v$ = 0.60-0.68.
Therefore I conclude that, although SU(3) 
symmetry is probably violated by polarized sea quark distributions,
a fit without contraints is quite unreliable. As I shall 
show in a moment, some bounds on polarized sea distributions must be 
taken into account.

Indeed, (\ref{ineq3p}), together with the well-known relations
\begin{eqnarray}
\Delta u + \Delta {\bar u} - \Delta d - \Delta {\bar d} = {\tilde a}_3,
\label{rel3}
\\
\Delta u + \Delta {\bar u}+ \Delta d + \Delta {\bar d}- 2\Delta s
-2\Delta {\bar s}= {\tilde a}_8, \label{rel8}
\end{eqnarray}
allows to deduce an important inequality about longitudinally 
polarized sea distributions. In eqs. (\ref{rel3}) and (\ref{rel8}) 
$\Delta q$  and $\Delta {\bar q}$ denote the first moments of the quark and 
antiquark polarized distributions,  ${\tilde a}_3$ = $(1.2670 \pm 0.0035) - 
\alpha_s/\pi$ and ${\tilde a}_8$ = $(0.585\pm 0.025) -\alpha_s/\pi$. 
Considering the splitting $\Delta q$ = $\Delta q_v +\Delta q_s$ into valence 
and sea contributions, eqs. (\ref{rel3}) and (\ref{rel8}) yield
\begin{equation}
\Delta d_v = - 0.341\pm 0.014 + \Delta {\tilde s} -\Delta {\tilde d},
\end{equation} 
having set $\Delta {\tilde q}$ = $\Delta q_s + \Delta {\bar q}$. 
Therefore the first inequality (\ref{ineq4}) implies 
\begin{equation}
\Delta {\tilde s} -\Delta {\tilde d} > 0.008\pm 0.014, \label{in33}
\end{equation} 
indicating that the SU(3) flavor symmetry for the longitudinally polarized 
sea distributions is very unlikely, although it cannot be completely excluded.
This result confirms the analysis by other authors\cite{grsv}.

Lastly, since $\Delta {\tilde s}$ is negative\cite{ls}, inequality (\ref{in33}) 
implies  
\begin{equation}
\Delta {\tilde d} < \Delta {\tilde s} < 0, \label {in99} 
\end{equation}
{\it i. e.}, a negatively polarized sea. This confirms the result that can be 
inferred\cite{alt} from data of polarized deep inelastic 
scattering\cite{emc}.
{\it Vice versa}, inequality (\ref{in99}) contradicts the 
interpretation in terms of a positive sea polarization of recent data of 
semi-inclusive deep inelastic scattering\cite{ja,sts}.
  
\section{CONCLUSION}

Here I recall the main results illustrated in my talk.

1. I have proved bound (\ref{ineq3p}) for any $Q^2$-value.

2. This bound is especially important in setting limits to fit parameters
and in discriminating between correct and wrong models. In particular, I find 
that the predictions of some models agree with that bound; on the contrary, 
recent distributions, resulting from best fits, violate it.

3. A bound on the polarized sea distributions is deduced, which turns 
out to agree with some analyses\cite{alt} of the EMC effect, 
also known as "spin crisis"\cite{emc}.


\begin{thebibliography}{9}

\bibitem{lss} E. Leader, A.V. Sidorov and D.B. Stamenov, Eur. Phys. J. 
C {\bf 23} (2002) 479; Phys. Lett. B {\bf 445} (1998) 232

\bibitem{bt} J. Bartelski and S. Tatur, Phys. Rev. D {\bf 65} (2002)
034002; hep-ph/0205089

\bibitem{rm} G.P. Ramsey, hep-ph/0211004, talk given at SPIN 2002, 
BNL, Sept. 9-13, 2002

\bibitem{so} J. Soffer, Phys. Rev. Lett. {\bf 74} (1995) 1292

\bibitem{ma} B.-Q. Ma, J. Phys. G: Nuc. Part. Phys. {\bf 17} (1991) L53

\bibitem{ss} I. Schmidt and J. Soffer,  Phys. Lett. B {\bf 407} (1997) 331

\bibitem{co} P.L. Chung et al., Phys. Rev. C {\bf 37} (1988) 37

\bibitem{mt}P.J. Mulders and R.D. Tangerman, Nucl. Phys. B {\bf 461} (1996) 197

\bibitem{pp} G. Parisi and R. Petronzio, Phys. Lett. B {\bf 62} (1976) 331

\bibitem{gr} M. Glueck and E. Reja, Nucl. Phys. B {\bf 130} (1977) 76

\bibitem{jr} R.L. Jaffe and G.G. Ross, Phys. Lett. B {\bf 93} (1980) 313

\bibitem{bdr} V. Barone, A. Drago and P. Ratcliffe, Phys. Rept. {\bf 359} 
(2002) 1


\bibitem{jj} R.L. Jaffe and X. Ji, Phys. Rev. Lett. {\bf 67} (1991) 552;
Nucl. Phys. B {\bf 375} (1992) 527

\bibitem{sv} S. Scopetta and V. Vento, Phys. Lett. B {\bf 424} (1997) 31

\bibitem{sw} K. Suzuki and W. Weise, Nucl. Phys. A {\bf 634} (1998) 141
 
\bibitem{mg} A. Manohar and H. Georgi, Nucl. Phys. B {\bf 234} (1984)
189

\bibitem{ck} R. Carlitz and J. Kaur, Phys. Rev. Lett. {\bf 38} (1977) 673

\bibitem{bst} C. Bourrely, J. Soffer and O. Teryaev, Phys. Lett. B 
{\bf 420} (1998) 375

\bibitem{blt} C. Bourrely, E. Leader and O. Teryaev, hep-ph/9803238, 
Talk given at the VII Workshop on High energy Spin Physics (SPIN-97),
Dubna, July 7-12, 1997

\bibitem{ar} G. Altarelli, S. Forte and G. Ridolfi, Nucl. Phys. B {\bf 534}
(1998) 277; S. Forte, M. Mangano and G. Ridolfi, Nucl. Phys. B {\bf 602} 
(2001) 585   

\bibitem{bf} R. Ball  and  S. Forte, Nucl. Phys. B {\bf 425} (1994) 
516 

\bibitem{dis3} E. Di Salvo, Il Nuovo Cimento A {\bf 111} (1998) 539

\bibitem{l1} S. Kretzer, E. Leader and E. Christova, Acta Phys. 
Polon. B {\bf 33} (2002) 3743

\bibitem{is} N. Isgur, Phys. Rev. D {\bf 59} (1999) 034013

\bibitem{grsv} M. Glueck, E. Reya, M. Stratmann and W. Vogelsang, 
Phys. Rev. D {\bf 63} (2001) 094005 

\bibitem{ls} E. Leader and D.B. Stamenov, Phys. Rev. D {\bf 67} (2003) 
037503 

\bibitem{alt} G. Altarelli and G. G. Ross, Phys. Lett. B {\bf 212} 
(1988) 391

\bibitem{emc} EMC collaboration, J. Ashman et al., Phys. Lett. B {\bf 206} 
(1988) 364; Nucl. Phys. B {\bf 328} (1989) 1 
 
\bibitem{ja} H.E. Jackson, Int. J. Mod. Phys. A {\bf 17} (2002) 3551

\bibitem{sts} U. Stosslein, Acta Phys. Polonica B {\bf 33} (2002) 2813
 
\end{thebibliography}
\end{document}